\documentclass[final,3p,times,twocolumn]{elsarticle}
\pdfoutput=1

\usepackage{amssymb}

\usepackage{geometry} 
\usepackage{graphicx}  
\usepackage{txfonts}   
\usepackage[colorlinks,linkcolor=blue,anchorcolor=blue,citecolor=blue,urlcolor=blue]{hyperref}
\usepackage{bm}
\usepackage[mathlines]{lineno}
\usepackage{multirow}
\usepackage[flushleft]{threeparttable}
\usepackage{color,soul}
\newcommand{\modR}[1] {{#1}}
\newcommand{\modB}[1] {{#1}}
\usepackage{natbib}   
\biboptions{sort&compress}

\journal{Carbon}
\begin{document}


\begin{frontmatter}



\title{
       Superior Hardness and Stiffness of Diamond Nanoparticles
       }



\author[1]{Alexander~Quandt}%

\author[2]{Igor~Popov}%

\author[3]{David~Tom\'{a}nek%
\corref{cor}} %
\ead{tomanek@msu.edu}

\address[1]{Mandelstam Institute for Theoretical Physics
            and School of Physics,
            University of the Witwatersrand,
            2050 Johannesburg, South Africa}

\address[2]{Institute of Physics Belgrade and
            Institute for Multidisciplinary Research,
            University of Belgrade,
            Belgrade, Serbia}

\address[3]{Physics and Astronomy Department,
            Michigan State University,
            East Lansing, Michigan 48824, USA}

\cortext[cor]{Corresponding author} %


\begin{abstract}
We introduce a computational approach to estimate the hardness and
stiffness of diamond surfaces and nanoparticles by studying their
elastic response to atomic nanoindentation. Results of our {\em ab
initio} density functional calculations explain the observed
hardness differences between different diamond surfaces and
suggest bond stiffening in bare and hydrogenated fragments of
cubic diamond and lonsdaleite. The increase in hardness and
stiffness can be traced back to bond length reduction especially
in bare nanoscale diamond clusters, a result of compression that
is driven by the dominant role of the surface tension.
\end{abstract}



\begin{keyword}
$\it{ab~initio}$, calculation, stability, hardness, nanoparticle,
DFT
\end{keyword}

\end{frontmatter}




\section{Introduction}
\modR{%
In the field of ultrahard materials, the role of diamond as the
hardest material on Earth seems to be well established. Not long
ago, this fact has been disputed by reports that compressed
fullerenes\cite{{Blank-Ultrahard-1998},{Wang-Long-Range-2012}},
nanotubes\cite{Popov-superhard-nanotube} and
graphite\cite{{Mao-Bonding-2003},{Irifune-Ultrahard-2003},
{Tanigaki-Observation-2013}}, which occur as highly disordered and
twinned nanocrystalline structures, may be still harder. %
} %
Also crystalline C$_3$N$_4$ was initially believed to be harder
than diamond due to its high bulk modulus~\cite{ALiu89}, but
ultimately turned out to be softer due to its inferior shear
modulus~\cite{{Teter96},{AQuandt13}}. %
On the macro-scale, mechanical hardness is commonly associated
with plastic deformations introduced by an external force, whereas
mechanical stiffness is associated with resistance to compression
and shear in the elastic regime. This distinction becomes blurred
on the nanometer scale, where the energy cost of introducing
plastic deformations exceeds that of fracture~\cite{Griffith1920}.
There, a scratch test appears to be a more suitable measure of
hardness, since the harder system need not undergo irreversible
plastic deformations.

So far, theoretical attempts to correlate mechanical hardness with
a particular bulk crystal structure have been mostly
disappointing\cite{{Chernozatonskii00},{DT169},%
{Li-Superhard-2009},{Umemoto-Body-Centered-2010},%
{Wang-Low-temperature-2011},%
{Selli-Superhard-2011},%
{Amsler-Crystal-2012},{Kvashnina-Investigation-2013}}. %
More recently, theoretical and experimental
studies have established a correlation between hardness, %
stiffness, and linear elastic constants in covalently bounded
materials~\cite{{Chen-hardness},{Mukhanov-hardness},{Jiang-hardness}}. %
\modR{ %
A new interesting evidence suggests that indentation hardness may
be proportional to the gravimetric density in carbon
materials.~\cite{Kang2020}. %
} %
Progress in computational materials science suggests that {\em ab
initio} calculations should be a valuable approach to determine
the stiffness and hardness of systems beyond the reach of common
experimental techniques. Observations in polycrystalline materials
including cubic boron nitride indicate an increasing hardness with
decreasing size of
the nanocrystallites%
~\cite{{Dubrovinskaia-Superhard-2007},{Solozhenko-Creation-2012}}.
We find it conceivable that also diamond
nanoparticles\cite{Dahl-Nat99} should be harder than their
macroscopic counterparts due to the dominant role of the surface
tension, which compresses the nanoparticles and stiffens the
interatomic interactions in the anharmonic regime. In the
macro-scale counterpart, stiffening of diamond under pressure is
evidenced in the non-vanishing third-order elastic
constants~\cite{Grimsditch78}.

At this point, we must emphasize that dislocation defects are
absent in nanosized particles due to the associated large energy
penalty. Consequently, Hall-Petch
strengthening~\cite{{VanSwygenhovenSci02},{Mo12}}
associated with dislocation motion does not occur in
nanoparticles. This fact sets nanostructures apart from their bulk
counterparts.

\begin{figure}[t]
\includegraphics[width=0.95\columnwidth]{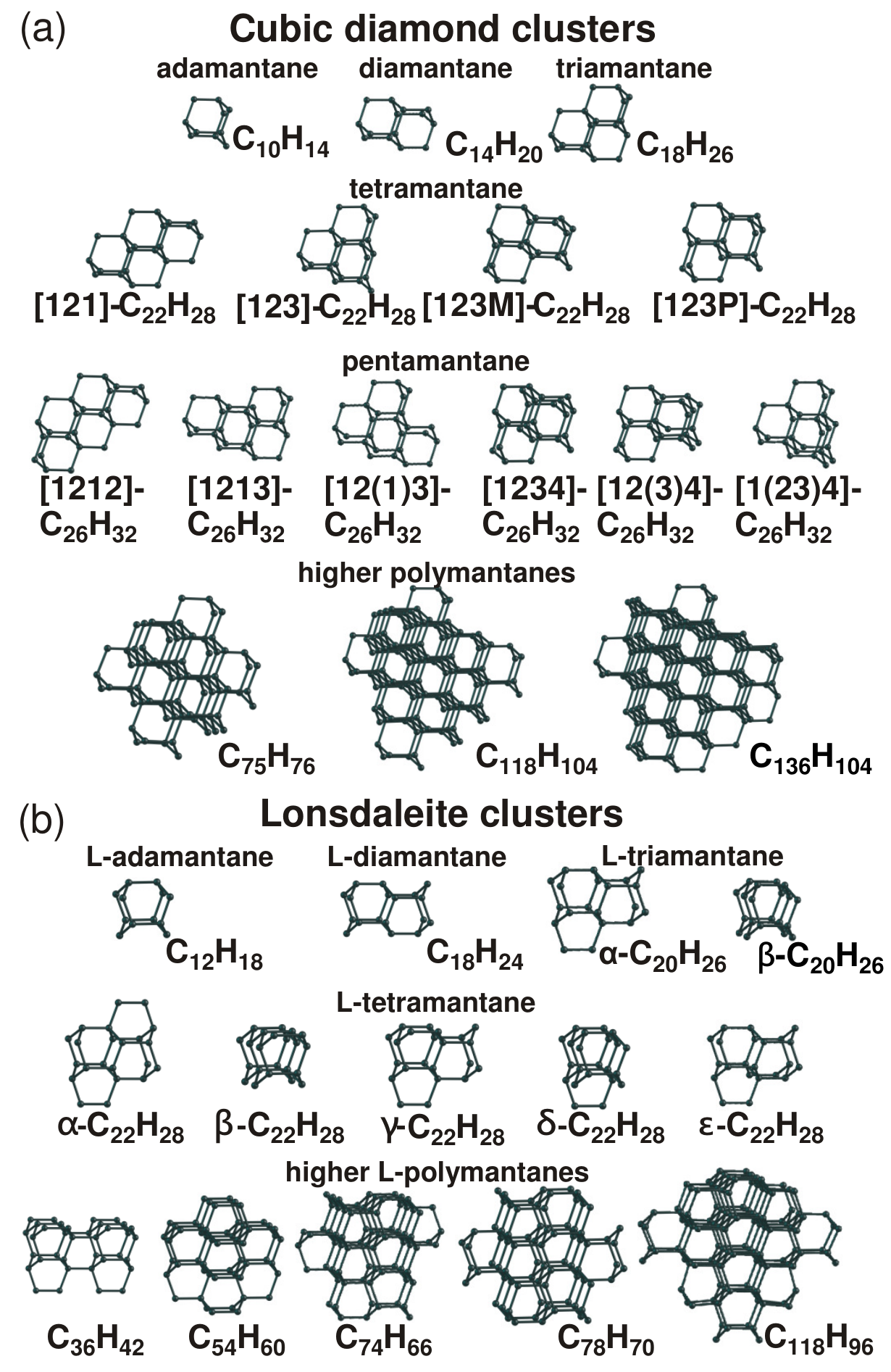}
\caption{Ball-and-stick models of (a) cubic diamond and (b)
lonsdaleite nanoparticles. Terminating hydrogen atoms
have been omitted for clarity.} %
\label{fig1}
\end{figure}



Here we introduce a computational approach to estimate and compare
the hardness and stiffness of diamond surfaces and nanoparticles,
independent of size, by studying their elastic response to atomic
nanoindentation. Results of our {\em ab initio} density functional
calculations explain the observed stiffness differences between
different diamond surfaces and indicate the occurrence of bond
stiffening in bare and hydrogenated fragments of cubic diamond and
of lonsdaleite. The increase in stiffness, especially in bare
diamond fragments, can be traced back to bond length reduction
that is driven by compression and caused by the surface tension.
In absence of plastic deformations on the nanometer scale,
increased stiffness corresponds to an increase in hardness.

\begin{figure}[t]
\includegraphics[width=1.0\columnwidth]{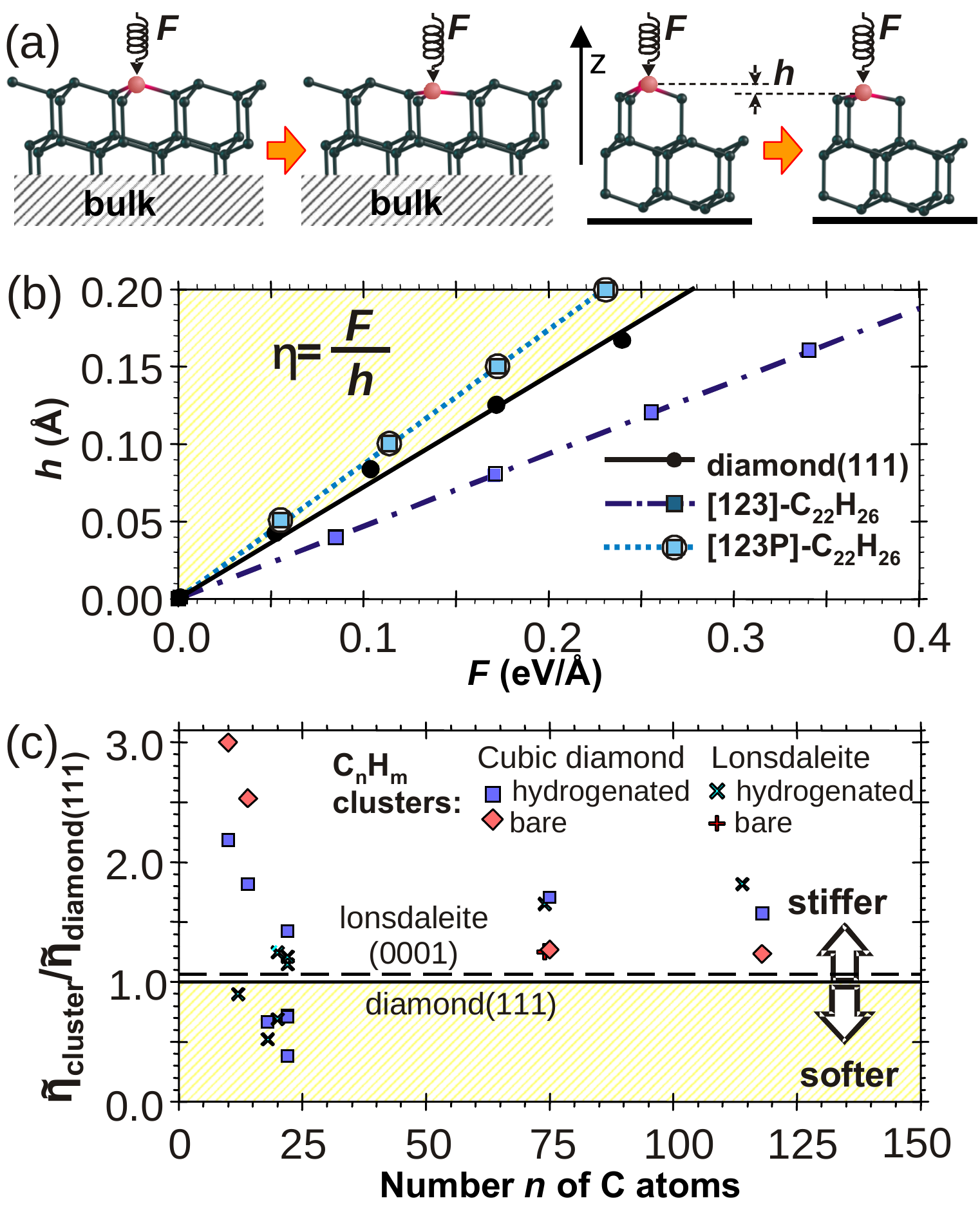}
\caption{%
\modB{ %
(a) Schematic of an atomic nanoindenter. The displacement $h$ of a
surface atom, highlighted in red, after being subject to force
$F$, serves as a local probe of of hardness and stiffness of
crystal surfaces and nanoparticles. Bulk atoms far underneath the
extended surface and at the bottom of nanoparticles are
constrained in the direction of the force. %
} %
(b) Calculated $F-h$ relationship at the (111) cubic diamond
surface and in two tetramantane isomers. %
(c) Normalized hardness $\tilde\eta$ in bare and hydrogen
terminated nanoparticles of cubic diamond and lonsdaleite compared
to the corresponding quantity for the (111) surface of cubic
diamond. %
Shaded regions in (b) and (c) represent response in systems softer
than diamond. } %
\label{fig2}
\end{figure}

\section{Results}

%

The relaxed geometries of hydrogen-terminated C$_n$H$_m$
diamondoid nanoparticles with $10{\le}n{\le}136$ carbon atoms,
obtained as fragments of cubic diamond and lonsdaleite, also
called hexagonal diamond, are shown in Fig.~\ref{fig1}. Since all
carbon atoms are $sp^3-$hybridized in these hydrogen terminated
systems, the equilibrium atomic arrangement is very close to that
in the bulk structure. The situation is very different in bare
carbon nanoparticles, where a significant fraction of surface
atoms with unsaturated bonds causes large-scale reconstruction of
the structures. The equilibrium structure of C$_n$
nanoparticles~\cite{DT049} comprises $sp^1-$bonded chains and
rings for $n<20$ and $sp^2-$bonded fullerenes for $n{\ge}20$. Some
small nanoparticles, including the C$_{10}$ adamantane and
C$_{14}$ diamantane, maintain their strained diamond-like
morphology as metastable structures. We found all larger
nanoparticles in this study, with diameters up to
${\approx}7$~{\AA}, to be unstable with respect to surface
graphitization due to the dominant role of unsaturated bonds at
the surface. This is true even in very large nanoparticles such as
C$_{136}$, where half the atoms change their hybridization from
$sp^3$ to $sp^2$.


\begin{table*}
\centering %
\caption{Observed hardness $H$ and calculated
         normalized hardness $\tilde\eta$ at the (111) and (100)
         surfaces of cubic diamond and the (0001) surface of
         lonsdaleite, as well as ratios of these quantities. }
\begin{threeparttable}
\begin{tabular}{|l|cc|c|c|c|}
\hline
& \multicolumn{2}{c|}{Diamond} & & {Lonsdaleite} & \\
& (111) & (100) & {\large$\frac{(100)}{(111)}$} %
  & (0001) & {\large$\frac{(0001)}{(111)}$} \\
\hline %
$H$~(GPa) %
& $167{\pm}5$\tnote{a} %
  & $137{\pm}6$\tnote{a} %
  & $0.82{\pm}0.06$ & -- & -- \\
& $117$\tnote{b} %
  & $95$\tnote{b} %
  & $0.81$ & & \\
$\tilde\eta$~(eV/{\AA}$^3$) %
  & 0.72 & 0.64 & 0.88 & 0.74 & 1.02 \\
\hline
\end{tabular}
\begin{tablenotes}
\item[a] Ref.~\cite{Blank-Ultrahard-1998} %
\item[b] Ref.~\cite{Richter-2000}
\end{tablenotes}
\end{threeparttable}
\label{table1}
\end{table*}


As mentioned in the Introduction, mechanical hardness $H$ is
commonly associated with resilience to plastic deformations
introduced by an external force. In macroscopic structures, $H$ is
measured by nanoindentation and defined by the ratio of the load
acting on a sharp nanoindenter and the resulting indentation
depth, as indicated schematically in Fig.~\ref{fig2}(a). It is an
integral characteristic of a solid that reflects resistance to
compression and shear and depends on quantities such as ductility,
elastic stiffness, plasticity, strength, toughness, and viscosity.
Since this complex response is hard to reproduce by {\em ab
initio} techniques, a number of empirical approaches have been
developed in recent years to estimate this
quantity\cite{Tian-Microscopic-2012}. Model
calculations\cite{{Simunek-Hardness-anisotropy},{Gao-Hardness-anisotropy}},
which have relied on simplified expressions based on a combination
of valence charges, bond ionicities and interatomic distances,
have so far failed to describe the dependence of hardness on the
surface orientation in extended solids. Due to their dependence on
a suitable choice of parameters, such model approaches are
typically limited to a specific class of systems.

In principle, direct calculations of
indentation~\cite{Richter-2000} should be able to describe
hardness anisotropy. Such studies would, however, necessitate very
large unit cells that currently exceed the scope of accurate
{\em ab initio} calculations. %
Here we introduce an alternative way to predict differences in
hardness based on the elastic response to particular deformations
that is based on {\em ab initio} total energy calculations. We
have considered specifically the Rockwell nanoindentation
technique~\cite{Gilman-2009}, which relates hardness to the
indentation depth caused by a conical nanoindenter that is rammed
into a surface by a given force. To extend our results to
nanoparticles, which are much smaller than any nanoindenter, we
have identified an individual surface atom as a nanoindenter.
Then, we relate the local hardness and stiffness to the $\eta=F/h$
ratio of the normal force $F$ to the atomic displacement $h$.

This approach naturally describes the response to compression and
shear on the atomic scale and allows us to discriminate between
different surface orientations. The atomic nanoindenter is shown
schematically in Fig.~\ref{fig2}(a) for semiinfinite surfaces and
nanoparticles. %
Our calculations determine directly the chemical stiffness of
interatomic bonds at the diamond surface. This quantity is related
to the earlier-defined chemical hardness~\cite{Gilman-2009}, which
measures the resistance to a change in chemical bonding. As
demonstrated earlier~\cite{Yang-1987}, the indentation hardness is
a monotonic function of the chemical hardness density. %

In our computational nanoindentation study, with a setup depicted
in Fig.~\ref{fig2}(a), we displace a particular atom by the
distance $h$ normal to the surface, relax the system, and
determine the force acting on the nanoindenter atom from
$F=-{\partial}E_{total}/{\partial}h$.
\modR{ %
This approach requires specific structural constraints, which we
specify in Section~\ref{CompTech} on Computational Techniques. %
} %

For indentation depths $h$ not exceeding a fraction of the
carbon-carbon bond length $d_{CC}$, we find a linear relationship
between $h$ and $F$, as seen in our results for the (111) surface
of cubic diamond and two tetramantane isomers in
Fig.~\ref{fig2}(b). Our results indicate that the force constant
$\eta$ of the $[123]$ diamond isomer is larger and that of the
$[123P]$ isomer is lower than that of the (111) surface of cubic
diamond. We conclude that the hardness of these particular diamond
fragments is close to, and may even exceed that of the bulk
crystalline diamond structure.

\begin{figure}[!tb]
\includegraphics[width=1.0\columnwidth]{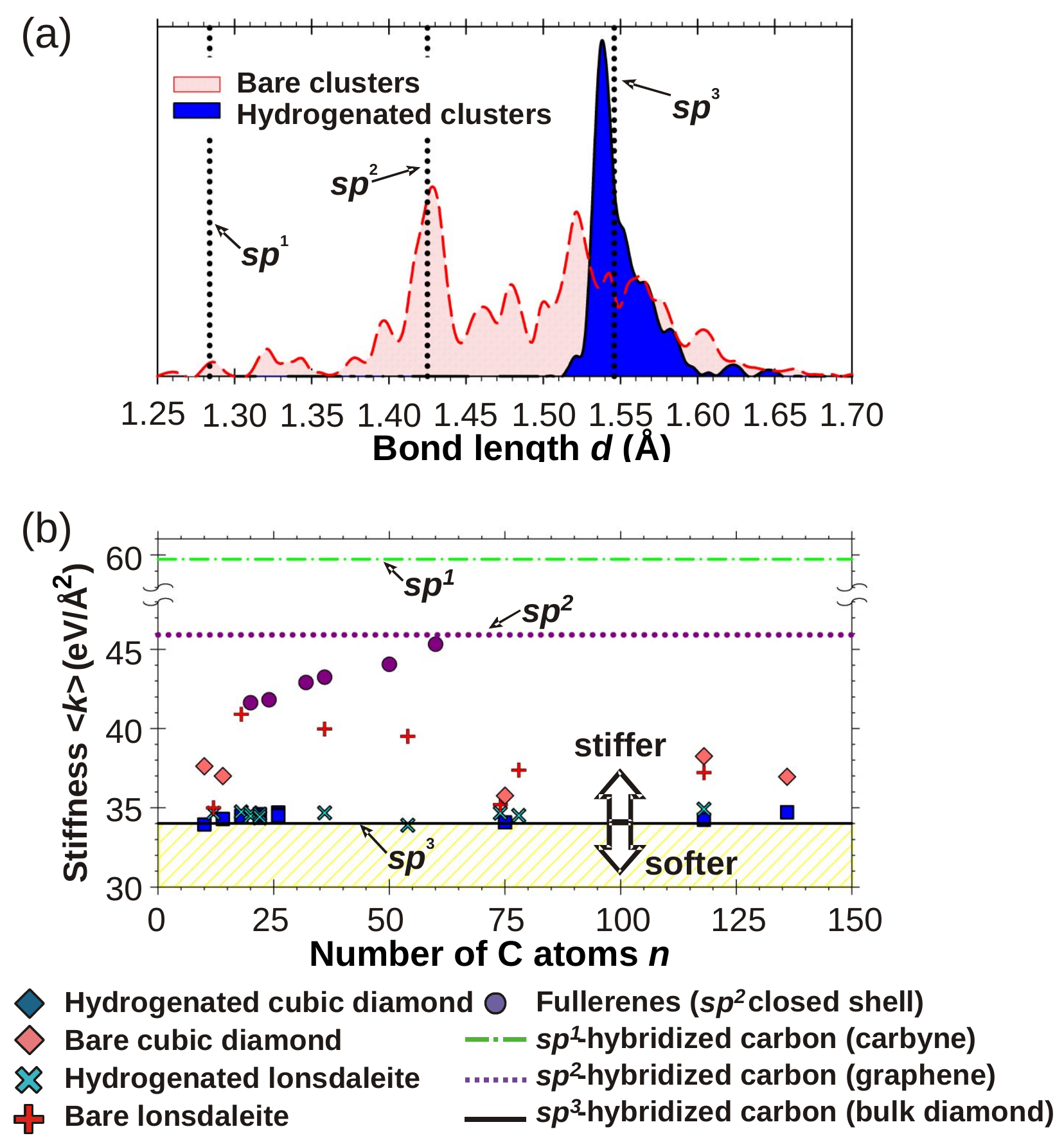}
\caption{%
(a) Bond length distribution in C$_n$H$_m$ diamond nanoparticles
of Fig.~\ref{fig1}. The light (red) shaded area below the dashed
line represents bare C$_n$ nanoparticles, and the dark (blue)
shaded area below the solid line represents
hydrogenated nanoparticles. %
(b) Comparison between the average bond stiffness
${\langle}k{\rangle}$, defined in Eq.~(\protect\ref{eq2}), in the
nanoparticles of panel (a) and in $sp^1$, $sp^2$ and $sp^3$
bonded bulk systems. %
\label{fig3}}
\end{figure}

To compare the hardness of different crystal surfaces, we
introduce the normalized hardness $\tilde\eta$, which we define by
\begin{equation}
\tilde\eta = \frac{\eta}{A} = \frac{F}{h{\cdot}A} \;, %
\label{eq1}
\end{equation}
where $A$ is the area per atom at a particular surface. We have
combined observed hardness values $H$ with our calculated values
of the related quantity $\tilde\eta$ for different surfaces of
cubic diamond and lonsdaleite in Table~\ref{table1}. Experimental
data indicate that the (111) surface is the hardest surface of
cubic diamond and that the (100) surface is 18\% softer. Even
though our computational approach does not provide absolute
hardness values, the calculated ratio
$\tilde\eta(100)/\tilde\eta(111)=0.88$ agrees well with the ratio
of the observed hardness values $H(100)/H(111)=0.82$ in cubic
diamond\cite{Blank-Ultrahard-1998}. There are no experimental
observations for the lonsdaleite structure, which occurs only as
inclusions in cubic diamond and is believed to be somewhat harder.
Based on our calculated values of $\tilde\eta$ listed in
Table~\ref{table1}, we believe that the (0001) surface of
lonsdaleite may be 2\% harder than the (111) surface of cubic
diamond.

Results in Table~\ref{table1} indicate that presence of
lonsdaleite alone may not explain reported hardness values that
are significantly higher than those of cubic
diamond\cite{{Blank-Ultrahard-1998},{Popov-superhard-nanotube},%
{Wang-Long-Range-2012},{Mao-Bonding-2003},{Irifune-Ultrahard-2003},%
{Tanigaki-Observation-2013}}. Therefore, we determine the
normalized hardness $\tilde\eta$ also for hydrogen-terminated and
bare diamond nanoparticles. For a reasonable comparison, we have
aligned each nanoparticle so that the topmost atom of the
unrelaxed structure, which will be subject to force $F$, belongs
to the (111) surface of cubic diamond or to the corresponding
(0001) surface of lonsdaleite. Since the area per atom is affected
by the net contraction of the nanoparticle, we estimated its
surface area $S$ from that of a polyhedron spanned by the nuclei
of the outermost atoms. We then used $A=A_i{\times}(S_f/S_i)$ for
the atom area in Eq.~(\ref{eq1}), where $A_i$ is the area per atom
at the corresponding infinite surface and $S_f/S_i$ is the ratio
of total nanoparticle surface areas in the final (f) and the
initial (i) structures.

We present our results for $\tilde\eta$ in hydrogen covered and
bare nanoparticles of cubic diamond as well as lonsdaleite in
Fig.~\ref{fig2}(c) and compare them to those for the (111) surface
of cubic diamond. Our results indicate that a significant fraction
of diamond nanoparticles appears to be significantly harder than
the hardest diamond surface. We find large differences in the
calculated values of $\tilde\eta$ even between different isomers
of the same nanostructure, such as the $[123]$ and $[123P]$
isomers of tetramantane. Their different elastic response,
depicted in Fig.~\ref{fig2}(b), reflects the simple fact that
particular structures may expand more or less easily in the plane
normal to the applied force. This flexibility is partly suppressed
in polycrystalline bulk assemblies of nanoparticles and also in
larger free-standing structures, which, however, approach bulk
diamond values with increasing nanoparticle size.

\section{Discussion}


There is an intuitive explanation for our finding that the
apparent hardness increases with decreasing size of diamond
nanoparticles. We need to note at this point that hardness
enhancement in nanoparticles of diamond and other solids is
fundamentally different from the behavior observed in
macro-structures, often described as the Hall-Petch effect, which
is associated with the nucleation and motion of
dislocations~\cite{{VanSwygenhovenSci02},{Mo12}}. As mentioned
earlier, plastic deformations do not occur in nanoparticles due to
the associated high energy cost.

In nanoparticles with a significant portion of surface atoms,
surface tension reduces significantly the surface area and thus
the interatomic bond length $d_{CC}$. We have determined the bond
length distribution in all nanoparticles presented in
Fig.~\ref{fig1} and plot this quantity separately for bare and for
hydrogenated nanoparticles in Fig.~\ref{fig3}(a). Termination by
hydrogen reduces the surface energy and provides a bulk-like
bonding environment even for carbon atoms at the surface.
Therefore, bond lengths in hydrogenated nanoparticles are all
close to the 1.54~{\AA} value found in $sp^3-$hybridized diamond.
On the other hand, we observe a significant bond length
contraction in bare diamond nanoparticles. %
\modR{ %
Under-coordinated atoms at the surface, which dominate in small
nanoparticles, reconstruct to form a ``net'' that contains and
compresses the ``bulk'' of the structure in a ``snug fit''. Most
surface atoms relax to a more favorable $sp^2-$like graphitic
bonding geometry with $d_{CC}{\approx}1.42$~{\AA}. Only a small
fraction of two-fold coordinated surface atoms is bonded in an
$sp^1-$like carbyne environment with $d_{CC}{\approx}1.28$~{\AA}. %
} %
The degree of bond contraction we find in nano-diamond agrees with
estimates for nanometer-sized diamond particles based on elastic
constants and surface energy. Due to the anharmonicity in the
interatomic bonds, the net bond contraction should cause a
stiffening in particular of bare
nanoparticles. %
\modR{ %
This reasoning is consistent with the recent
observation~\cite{Kang2020} that indentation hardness increases at
higher densities caused by bond contraction and reconstruction
in nanostructures. %
} %



To validate our interpretation, we estimated the bond stiffness in
nanoparticles considered in our study. In each optimized
nanoparticle, we first determined the average bond length
${\langle}d_{CC,0}{\rangle}$ and the bond energy
$E_{b,0}=E_{coh,0}/N_b$ by dividing the cohesive energy $E_{coh}$
by the number of nearest-neighbor bonds $N_b$. We then uniformly
expanded or contracted the nanoparticle and determined the
corresponding average bond length ${\langle}d_{CC,0}{\rangle}$ and
bond energy $E_b$. Finally, we determined the average bond
stiffness ${\langle}k{\rangle}$ in a given particle using
\begin{equation}
|E_b-E_{b,0}| = \frac{1}{2} {\langle}k{\rangle} %
({\langle}d_{CC}{\rangle} - {\langle}d_{CC,0}{\rangle})^2 \;. %
\label{eq2}
\end{equation}
We plot the quantity ${\langle}k{\rangle}$ for C$_n$H$_m$
nanoparticles and compare it to that of $sp^3$, $sp^2$ and $sp^1$
hybridized systems in Fig.~\ref{fig3}(b). Our results indicate
that the bond stiffness in hydrogenated nanoparticles is
comparable to that in $sp^3-$hybridized diamond and is typically
much higher in bare nanoparticles, approaching the higher bond
stiffness of $sp^2-$hybridized graphene. For the sake of fair
comparison, we also present bond stiffness values of fullerenes in
Fig.~\ref{fig3}(b). These hollow graphitic nanoparticles display
$sp^2$ bonding with a small $sp^3$ admixture, and their bond
stiffness values are in the expected range. We conclude that the
enhanced bond stiffness in bare nanoparticles is caused by surface
reconstruction from dominant $sp^3$ to at least partial
$sp^2-$type bonding. We may expect that diamond nanoparticles with
a graphitized outer surface, which are often observed
experimentally\cite{Welz-Nucleation-2003}, may have stiffer bonds
than diamond.

At this point we should re-emphasize that the connection between
bond stiffness, reflecting elastic response, and hardness, which
describes irreversible plastic deformations, is only indirect.
Bond stiffness describes the resistance of bonds to stretching and
compression, which we model by uniformly compressing the entire
structure. On the other hand, hardness characterizes the
resistance of a structure to indentation, which we model by
displacing one single surface atom. The hardest nanoparticles in
our study were bare and hydrogen-terminated C$_{10}$H$_x$
adamantane and C$_{14}$H$_x$ diamantane nanoparticles. With the
exception of these two systems, hydrogenated nanoparticles were
found to be harder than bare nanoparticles. The significant
increase in nominal hardness, which we found in ultra-small
nanoparticles, diminishes rapidly in systems containing hundreds
of carbon atoms and approach rapidly well-established bulk values.


\section{Summary and Conclusions}


In conclusion, we have introduced a computational approach to
estimate and compare the hardness and stiffness of both
single-crystal surfaces and nanoparticles of diamond, which are
too small for indentation experiments, by studying their elastic
response to atomic nanoindentation. Results of our {\em ab initio}
density functional calculations of this nanoindentation process
correlated well with the observed differences in hardness between
different diamond surfaces. More important, we find bond
stiffening in bare and hydrogenated fragments of both cubic
diamond and lonsdaleite. The increase in stiffness, especially in
bare nano-diamond particles, can be traced back to bond length
reduction. The net average bond compression is driven by the
dominant role of the surface tension and leads to surface
reconstruction. Since plastic deformations do not occur on the
nanometer scale, increased stiffness indicates an increase in
hardness. It is likely that the scratch hardness of diamond
nanoparticles, which are used to cover drill heads, may exceed
that of monocrystalline diamond.

\section{Computational Techniques}
\label{CompTech}

\modR{ %
\subsection{Representation of Nanoindentation}
} %

\modR{ %
We represent a periodic infinite surface, shown in the left panels
of Fig.~\ref{fig2}(a), by a unit cell with a finite surface area,
which contains infinitely many atoms below the surface. In
principle, the displacement of the atomic nanoindenter along the
$-z$ direction into the surface may cause all atoms within the
unit cell to move, but the shape of the unit cell will not change.
In our study, we only allow atomic displacement within a thick
surface region above a frozen bulk structure that balances the
force caused by the nanoindenter. $E_{total}$ is determined
for the optimized geometry. %
} %

\modR{ %
In finite nanoparticles, depicted in the right panels of
Fig.~\ref{fig2}(a), there are no symmetry restrictions on atomic
displacements or global shape deformations. To provide a realistic
description of nanoindentation in a previously relaxed
nanoparticle, we first displace the atomic nanoindenter along the
$-z$ direction. Next, we fix the $z$-coordinates, but not the $x$-
and $y$- coordinates, of all bottom atoms of the nanoparticle to
balance the force caused by the nanoindenter. Finally, we relax
all remaining atomic degrees of freedom and determine $E_{total}$.
For both infinite surfaces and finite nanoparticles, we obtain the
force that caused the deformation from
$F=-{\partial}E_{total}/{\partial}h$. %
} %

\modR{ %
\subsection{Total Energy Formalism}
} %

Our calculations of the optimum atomic structure, stability and
elastic properties of diamond surfaces and nanoparticles are based
on the density functional theory
(DFT)\cite{{Hohenberg-Inhomogeneous-1964},
{Kohn-Self-Consistent-1965}}. We used the PBE-PAW approximation
\cite{Perdew-Generalized-1996} to DFT, as implemented in the
\textsc{VASP}%
\cite{{Kresse-Ab-1993},{Kresse-Ab-1994},{Kresse-Efficient-1996}}
code. All systems have been represented using periodic boundary
conditions and a plane-wave energy cutoff of 520~eV. %
\modR{ %
Spurious interaction between neighboring particles has been
suppressed by requiring the closest-approach distance between
adjacent surfaces to exceed~\cite{cutoff-distance} $6$~{\AA}. %
} %
All structures have been relaxed until all forces acting on
atoms were less than 0.01~eV/{\AA}.


{\noindent\bf Author Information}\\

{\noindent\bf Corresponding Author}\\
$^*$E-mail: {\tt tomanek@msu.edu}

{\noindent\bf Notes}\\
The authors declare no competing financial interest.


\section*{Acknowledgments}

We thank Mikhail Yu.~Popov and Arthur G.~Every for useful
discussions. This study was supported by the NSF/AFOSR EFRI 2-DARE
grant number EFMA-1433459.
Computations have been performed by Liubov Yu.~Antipina and Pavel
B.~Sorokin while visiting Michigan State University in 2013-2014.
Computational resources have been provided by the Michigan State
University High Performance Computing Center and the Moscow State
University Supercomputer. AQ thanks the Materials for Energy
Research Group (MERG) and the DST-NRF Centre of Excellence in
Strong Materials (CoE-SM) at the University of the Witwatersrand
for support. AQ and DT also thank the Mandelstam Institute for
Theoretical Physics (MITP) and the Simons Foundation, award number
509116, for support.


\begin{thebibliography}{10}
\expandafter\ifx\csname url\endcsname\relax
  \def\url#1{\texttt{#1}}\fi
\expandafter\ifx\csname
urlprefix\endcsname\relax\def\urlprefix{URL }\fi
\expandafter\ifx\csname href\endcsname\relax
  \def\href#1#2{#2} \def\path#1{#1}\fi

\bibitem{Blank-Ultrahard-1998}
V.~Blank, M.~Popov, G.~Pivovarov, N.~Lvova, K.~Gogolinsky,
V.~Reshetov,
  Ultrahard and superhard phases of fullerite {C}$_{60}$: Comparison with
  diamond on hardness and wear, Diamond Relat. Mater. 7~(2--5) (1998) 427--431.

\bibitem{Wang-Long-Range-2012}
L.~Wang, B.~Liu, H.~Li, W.~Yang, Y.~Ding, S.~V. Sinogeikin,
Y.~Meng, Z.~Liu,
  X.~C. Zeng, W.~L. Mao, Long-range ordered carbon clusters: A crystalline
  material with amorphous building blocks, Science 337~(6096) (2012) 825--828.

\bibitem{Popov-superhard-nanotube}
M.~Popov, M.~Kyotani, Y.~Koga, Superhard phase of single-wall
carbon nanotube,
  Physica B: Cond. Matt. 323~(1-4) (2002) 262--264.

\bibitem{Mao-Bonding-2003}
W.~L. Mao, H.~Mao, P.~J. Eng, T.~P. Trainor, M.~Newville, C.~Kao,
D.~L. Heinz,
  J.~Shu, Y.~Meng, R.~J. Hemley, Bonding changes in compressed superhard
  graphite, Science 302~(5644) (2003) 425--427.

\bibitem{Irifune-Ultrahard-2003}
T.~Irifune, A.~Kurio, S.~Sakamoto, T.~Inoue, H.~Sumiya, Ultrahard
  polycrystalline diamond from graphite, Nature 421 (2003) 599--600.

\bibitem{Tanigaki-Observation-2013}
K.~Tanigaki, H.~Ogi, H.~Sumiya, K.~Kusakabe, N.~Nakamura,
M.~Hirao,
  H.~Ledbetter, Observation of higher stiffness in nanopolycrystal diamond than
  monocrystal diamond, Nature Comm. 4 (2013) 2343.

\bibitem{ALiu89}
A.~Y. Liu, M.~L. Cohen, Prediction of new low compressibility
solids, Science
  245~(4920) (1989) 841--842.

\bibitem{Teter96}
D.~M. Teter, R.~J. Hemley, Low-compressibility carbon nitrides,
Science
  271~(5245) (1996) 53--55.

\bibitem{AQuandt13}
G.~S. Manyali, R.~Warmbier, A.~Quandt, J.~E. Lowther, {\it{Ab
initio}} study of
  elastic properties of super hard and graphitic structures of {C}$_3${N}$_4$,
  Comput. Mater. Sci. 69 (2013) 299--303.

\bibitem{Griffith1920}
A.~Griffith, {VI.} {The} phenomena of rupture and flow in solids,
Phil. Trans.
  Roy. Soc. (London) -- A 221 (1920) 163--198.

\bibitem{Chernozatonskii00}
L.~Chernozatonskii, N.~Serebryanaya, B.~Mavrin, The superhard
crystalline
  three-dimensional polymerized {C}$_{60}$ phase, Chem. Phys. Lett. 316~(3-4)
  (2000) 199.

\bibitem{DT169}
S.~Berber, E.~Osawa, D.~Tomanek, Rigid crystalline phases of
polymerized
  fullerenes, Phys. Rev. B 70~(8) (2004) 085417.

\bibitem{Li-Superhard-2009}
Q.~Li, Y.~Ma, A.~R. Oganov, H.~Wang, H.~Wang, Y.~Xu, T.~Cui, H.-K.
Mao, G.~Zou,
  Superhard monoclinic polymorph of carbon, Phys. Rev. Lett. 102~(17) (2009)
  175506.

\bibitem{Umemoto-Body-Centered-2010}
K.~Umemoto, R.~M. Wentzcovitch, S.~Saito, T.~Miyake, Body-centered
tetragonal
  {C}$_4$: $sp^3$ carbon allotrope, Phys. Rev. Lett. 104~(12) (2010) 125504.

\bibitem{Wang-Low-temperature-2011}
J.-T. Wang, C.~Chen, Y.~Kawazoe, Low-temperature phase
transformation from
  graphite to $sp^3$ orthorhombic carbon, Phys. Rev. Lett. 106~(7) (2011)
  075501.

\bibitem{Selli-Superhard-2011}
D.~Selli, I.~A. Baburin,
R.~Marto\ifmmode\check{n}\else\v{n}\fi{}\'ak,
  S.~Leoni, Superhard $sp^3$ carbon allotropes with odd and even ring
  topologies, Phys. Rev. B 84 (2011) 161411.

\bibitem{Amsler-Crystal-2012}
M.~Amsler, J.~A. Flores-Livas, L.~Lehtovaara, F.~Balima, S.~A.
Ghasemi,
  D.~Machon, S.~Pailh\`es, A.~Willand, D.~Caliste, S.~Botti, A.~San~Miguel,
  S.~Goedecker, M.~A.~L. Marques, Crystal structure of cold compressed
  graphite, Phys. Rev. Lett. 108 (2012) 065501.

\bibitem{Kvashnina-Investigation-2013}
Y.~A. Kvashnina, A.~G. Kvashnin, P.~B. Sorokin, Investigation of
new superhard
  carbon allotropes with promising electronic properties, J. Appl. Phys.
  114~(18) (2013) 183708.

\bibitem{Chen-hardness}
X.-Q. Chen, H.~Niu, D.~Li, Y.~Li, Modeling hardness of
polycrystalline
  materials and bulk metallic glasses, Intermetallics 19~(9) (2011) 1275--1281.

\bibitem{Mukhanov-hardness}
V.~A. Mukhanov, O.~O. Kurakevych, V.~L. Solozhenko, Thermodynamic
aspects of
  materials' hardness: prediction of novel superhard high-pressure phases, High
  Pressure Research 28~(4) (2008) 531--537.

\bibitem{Jiang-hardness}
X.~Jiang, J.~Zhao, X.~Jiang, Correlation between hardness and
elastic moduli of
  the covalent crystals, Computational Materials Science 50~(7) (2011)
  2287--2290.

\bibitem{Kang2020}
\modR{S.~Kang, Z.~Xiang, H.~Mu, Y.~Cai, Mechanical properties,
lattice thermal
  conductivity, infrared and {Raman} spectrum of the fullerite {C}$_{24}$,
  Phys. Lett. A 384~(1) (2020) 126035.}

\bibitem{Dubrovinskaia-Superhard-2007}
N.~Dubrovinskaia, V.~L. Solozhenko, N.~Miyajima, V.~Dmitriev,
O.~O. Kurakevych,
  L.~Dubrovinsky, Superhard nanocomposite of dense polymorphs of boron nitride:
  Noncarbon material has reached diamond hardness, Appl. Phys. Lett. 90~(10)
  (2007) 101912.

\bibitem{Solozhenko-Creation-2012}
V.~L. Solozhenko, O.~O. Kurakevych, Y.~L. Godec, Creation of
nanostuctures by
  extreme conditions high-pressure synthesis of ultrahard nanocrystalline cubic
  boron nitride, Adv. Mater. 24~(12) (2012) 1540--1544.

\bibitem{Dahl-Nat99}
J.~E. Dahl, J.~M. Moldowan, K.~E. Peters, G.~E. Claypool, M.~A.
Rooney, G.~E.
  Michael, M.~R. Mello, M.~L. Kohnen, Diamonxdoid hydrocarbons as indicators of
  natural oil cracking, Nature 399~(7) (1999) 54--57.

\bibitem{Grimsditch78}
M.~H. Grimsditch, E.~Anastassakis, M.~Cardona, Effect of uniaxial
stress on the
  zone-center optical phonon of diamond, Phys. Rev. B 18 (1978) 901--904.

\bibitem{VanSwygenhovenSci02}
H.~Van~Swygenhoven, Grain boundaries and dislocations, Science
296~(5565)
  (2002) 66--67.

\bibitem{Mo12}
Y.~Mo, D.~Stone, I.~Szlufarska, Strength of ultrananocrystalline
diamond
  controlled by friction of buried interfaces, J. Phys. D: Appl. Phys. 44~(40)
  (2011) 405401, and {\bf 45}, 069501 (2012) (E).

\bibitem{DT049}
D.~Tomanek, M.~A. Schluter, Growth regimes of carbon clusters,
Phys. Rev. Lett.
  67~(17) (1991) 2331--2334.

\bibitem{Richter-2000}
A.~Richter, R.~Ries, R.~Smith, M.~Henkel, B.~Wolf, Nanoindentation
of diamond,
  graphite and fullerene films, Diam. Rel. Mater. 9~(2) (2000) 170--184.

\bibitem{Tian-Microscopic-2012}
Y.~Tian, B.~Xu, Z.~Zhao, Microscopic theory of hardness and design
of novel
  superhard crystals, Int. J. Refract. Met. Hard Mater. 33 (2012) 93--106.

\bibitem{Simunek-Hardness-anisotropy}
A.~\ifmmode \check{S}\else \v{S}\fi{}im\ifmmode~\mathring{u}\else
  \r{u}\fi{}nek, Anisotropy of hardness from first principles: The cases of
  {ReB}$_{2}$ and {OsB}$_{2}$, Phys. Rev. B 80 (2009) 060103.

\bibitem{Gao-Hardness-anisotropy}
F.~Gao, Theoretical model of hardness anisotropy in brittle
materials, J. Appl.
  Phys. 112~(2) (2012) 023506.

\bibitem{Gilman-2009}
J.~Gilman, Chemistry and physics of mechanical hardness, John
Wiley \& Sons,
  2009.

\bibitem{Yang-1987}
W.~Yang, R.~G. Parr, L.~Uytterhoeven, New relation between
hardness and
  compressibility of minerals, Phys. Chem. Minerals 15~(2) (1987) 191--195.

\bibitem{Welz-Nucleation-2003}
S.~Welz, Y.~Gogotsi, M.~J. McNallan, Nucleation, growth, and
graphitization of
  diamond nanocrystals during chlorination of carbides, J. Appl. Phys. 93~(7)
  (2003) 4207.

\bibitem{Hohenberg-Inhomogeneous-1964}
P.~Hohenberg, W.~Kohn, Inhomogeneous electron gas, Phys. Rev.
136~(3B) (1964)
  864--871.

\bibitem{Kohn-Self-Consistent-1965}
W.~Kohn, L.~J. Sham, Self-consistent equations including exchange
and
  correlation effects, Phys. Rev. 140~(4A) (1965) 1133--1138.

\bibitem{Perdew-Generalized-1996}
J.~P. Perdew, K.~Burke, M.~Ernzerhof, Generalized gradient
approximation made
  simple, Phys. Rev. Lett. 77~(18) (1996) 3865--3868.

\bibitem{Kresse-Ab-1993}
G.~Kresse, J.~Hafner, Ab initio molecular dynamics for liquid
metals, Phys.
  Rev. B 47~(1) (1993) 558--561.

\bibitem{Kresse-Ab-1994}
G.~Kresse, J.~Hafner, Ab initio molecular-dynamics simulation of
the
  liquid-metal-amorphous-semiconductor transition in germanium, Phys. Rev. B
  49~(20) (1994) 14251--14269.

\bibitem{Kresse-Efficient-1996}
G.~Kresse, J.~Furthm\"uller, Efficient iterative schemes for ab
initio
  total-energy calculations using a plane-wave basis set, Phys. Rev. B 54
  (1996) 11169--11186.

\bibitem{cutoff-distance}
\modR{On a per-surface-atom basis, the interaction energy between
      diamond surfaces separated by $6$~{\AA} differs by less than
      $0.02$~meV from that for surfaces separated by $10$~{\AA}.}

\end{thebibliography}

\end{document}